\documentclass[a4paper]{article}

\usepackage{18lomcon}        
\usepackage{cite}             
\usepackage{epsfig}           
\usepackage{epstopdf}

\begin{document}


\title{THE MUON $g$-$2$ EXPERIMENT AT FERMILAB}

\author{Wesley Gohn \email{gohn@pa.uky.edu}
        for the Muon $g$-$2$ Collaboration
        }

\affiliation{University of Kentucky, Lexington, KY, USA}


\date{}
\maketitle


\begin{abstract}
  A new measurement of the anomalous magnetic moment of the muon, $a_{\mu} \equiv (g-2)/2$, will be performed at the Fermi National Accelerator Laboratory with data taking beginning in 2017. The most recent measurement, performed at Brookhaven National Laboratory (BNL) and completed in 2001, shows a 3.5 standard deviation discrepancy with the standard model value of $a_\mu$. The new measurement will accumulate 21 times the BNL statistics using upgraded magnet, detector, and storage ring systems, enabling a measurement of $a_\mu$ to 140 ppb, a factor of 4 improvement in the uncertainty the previous measurement. This improvement in precision, combined with recent improvements in our understanding of the QCD contributions to the muon $g$-$2$, could provide a discrepancy from the standard model greater than 7$\sigma$ if the central value is the same as that measured by the BNL experiment, which would be a clear indication of new physics.
\end{abstract}

\section{Introduction}

The anomalous magnetic moment of the muon was last measured by the Brookhaven experiment E821 in 1999-2001, resulting in a 3.5$\sigma$ discrepancy with the Standard Model of particle physics~\cite{Bennett:2006fi}. The Muon g-2 experiment E989 at Fermilab will improve the precision on $a_{\mu}$ by a factor of four by performing the measurement using 21 times the statistics of E821~\cite{Grange:2015fou}. The new experiment will attempt to answer the question, is this discrepancy an indication of new physics beyond the standard model?

 The muon has a magnetic dipole moment of $\vec \mu = g\frac{q}{2m}\vec s$, with $g=2$ for a pointlike particle~\cite{Dirac:1928hu}.  The Standard Model predicts effects from QED, electroweak theory, and QCD, such that $g_{SM} = 2_{Dirac} + \mathcal{O}(10^{-3})_{QED} + \mathcal{O}(10^{-9})_{EW} + \mathcal{O}(10^{-7})_{QCD}$. If a discrepancy with the standard model value is found, beyond standard model contributions to $g$-$2$ could come from SUSY, dark photons, extra dimensions, or other new physics. Significant theoretical effort has gone into understanding these contributions including~\cite{Studenikin:1990xm, Kowalska:2017iqv,Davier:2010nc,Aoyama:2012wk,Gnendiger:2013pva}. The error on the theoretical value is expected to be reduced on the timescale of the new experiment, due to efforts to reduce the hadronic contributions to $a_\mu$ from lattice calculations~\cite{Blum:2017cer,Blum:2016lnc} and improved measurements of lepton scattering cross-sections. A measurement to this precision at the BNL central value could provide a better than $7\sigma$ discrepancy from the standard model, which would provide clear evidence of new physics.

\section{The Experiment}

To perform the experiment, we inject polarized muons into a magnetic storage ring. Muons will precess in the magnetic field, and we measure the frequency $\omega_a$ via the timing of muon decays to positrons using 24 electromagnetic calorimeters~\cite{Kaspar:2016ofv,Fienberg:2014kka}. Measurements of the precession frequency and magnetic field lead to $a_\mu$. The anomalous precession frequency $\omega_a = \omega_s - \omega_c$, where the Cyclotron frequency $\omega_c = \frac{e}{m\gamma}B$ and the spin precession frequency $\omega_s = \frac{e}{m\gamma}B(1+\gamma a_\mu)$. 

To reduce the effect of electric fields, the muons are injected at a magic momentum with $\gamma = 29.3$, which cancels the second term in Eq.~\ref{eq:omega}. 

\begin{equation}
\vec\omega_a = -\frac{Qe}{m}[a_\mu \vec B - (a_\mu - (\frac{mc}{p})^2)\frac{\vec \beta \times \vec E}{c}],
\label{eq:omega}
\end{equation}

\noindent which leaves 

\begin{equation}
\vec\omega_a = -\frac{Qe}{m}a_\mu \vec B.
\label{eq:omegaa}
\end{equation}

The proton precession frequency $\omega_p$ is measured as a proxy for $\vec B$ leading to an expression for the anomalous moment as
\begin{equation}
 a_\mu = \frac{\omega_a / \omega_p}{\mu_\mu / \mu_p - \omega_a / \omega_p}.
\label{eq:omegap}
\end{equation}
The Measurement of $\omega_p$ is performed using NMR. Fixed NMR probes measure time variations of the field during data taking. A trolley with mounted NMR probes periodically circumnavigates the interior of the ring to perform precision measurements of the field in the muon storage region, performing 6000 magnetic field measurements per trolley run. Probes are calibrated to provide measurement to 35 ppb.

We plan to collect 21 times the BNL statistics, which will reduce our statistical uncertainty by a factor of four. To reduce systematic uncertainty, accelerator facilities will have $p_\pi$ closer to magic momentum, utilize a longer decay channel, and increase injection efficiency. Systematics on $\omega_a$ will be decreased from 180 ppb in E821 to 70 ppb by using an improved laser calibration, a segmented calorimeter, better collimator in the ring, and improved tracker. Systematics on $\omega_p$ will be decreased from 170 ppb in E821 to 70 ppb by improving the uniformity and monitoring of the magnetic field, increasing accuracy of position determination of trolly, better temperature stability of the magnet, and providing active feedback to external fields.

\begin{figure}[!t]
\centering
\begin{minipage}{5cm}
\centering
\includegraphics[width=5cm]{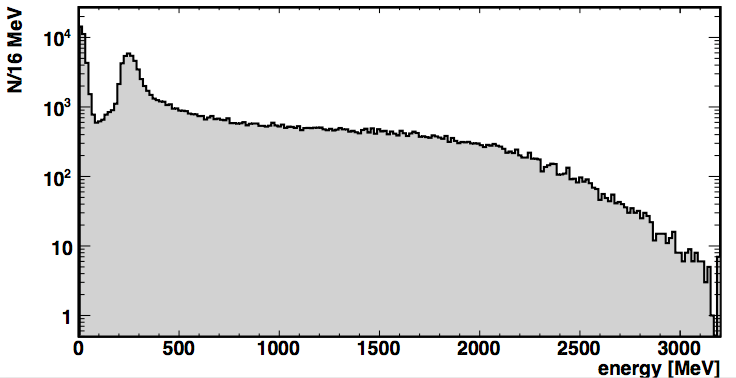}
\caption{Energy distribution from June 2017 data recorded in the calorimeter. The low energy peaks are from protons and lost muons.}
\end{minipage}
\hspace{1cm}
\begin{minipage}{5cm}
\centering
\includegraphics[width=5cm]{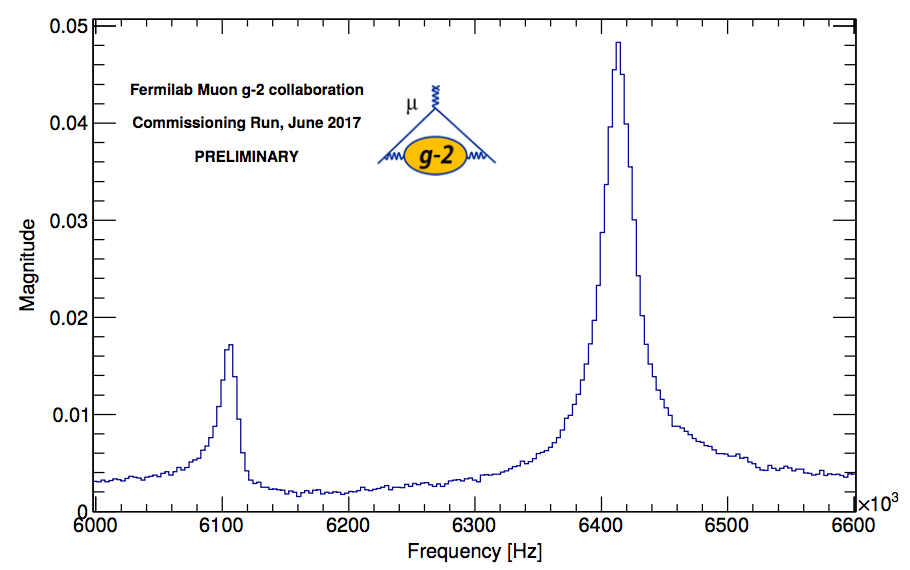}
\caption{Fourier Transform of data from fiber harps shows the proton cyclotron frequency and the betatron frequency of stored protons. }
\end{minipage}
\label{fig:detectors}
\end{figure}

\section{Current status}

\begin{figure}[!t]
\centering
\includegraphics[width=7cm]{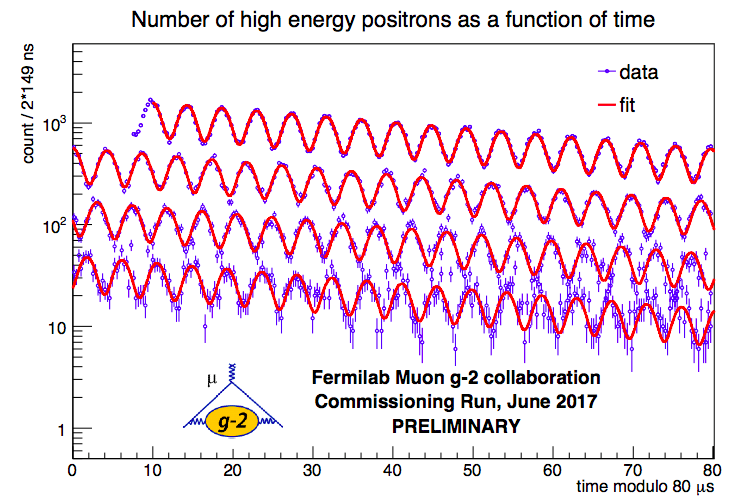}
\caption{Time distribution of muon decays from June 2017 engineering run.}
\label{fig:wiggle}
\end{figure}

Muon g-2 performed a 5 week engineering run during June-July 2017. The first beam was injected into the ring on May 31, 2017. The accelerator performed well during the run~\cite{Stratakis:2017uci}, though the beam was mostly protons with $\mathcal{O}(1\%)$ muons and had a fill rate of 0.1 Hz (compared nominal rate of 12 Hz expected during production operations). The effort of the engineering run was largely dedicated to optimizing muon injection into the ring by tuning the inflector magnet, electrostatic quadrupoles, and kickers to store the beam. Stored protons, muons, and positrons were detected with all 24 operational calorimeters, one of the three planned tracker stations, and the retractable fiber harp detectors, as shown in Fig.~\ref{fig:detectors}. A sufficient number of muon decays were observed during the run to see the muon precession signal, as shown in Fig.~\ref{fig:wiggle}, though the best precision from this data is expected to yield a result of no more than 50 ppm, which is two orders of magnitude below the E821 statistics.

\section{Conclusion}

The new Muon g-2 experiment at Fermilab will measure the anomalous magnetic moment of the muon to 4 $\times$ the precision of the previous BNL measurement. If the previously measured value holds, this could provide a 7$\sigma$ discrepancy from the standard model.
A successful commissioning run was performed in June of 2017 and the experiment will continue running this November. The 2017 run provided $10^{-5}$ of the required statistics, which is between the total statistics of the CERN II and CERN III measurements of $a_\mu$. The experiment's goal is for a BNL level (500 ppb) result from the 2018 data and the final 140 ppb measurement from data collected through 2020. 

\section*{Acknowledgments}

This work is supported by Fermi Research Alliance, LLC under Contract No. DE-AC02-07CH11359 with the United States Department of Energy.


%
%

\bibliography{lomcon}

\end{document}